\documentclass[12pt`]{article}

\newcommand  {\Ebar} {{\mbox{\rm$\mbox{I}\!\mbox{E}$}}}

\newsavebox{\zzzbar}
\sbox{\zzzbar}
  {\setlength{\unitlength}{0.9em}
  \begin{picture}(0.6,0.7)
  \thinlines
  \put(0,0){\line(1,0){0.6}}
  \put(0,0.75){\line(1,0){0.575}}
  \multiput(0,0)(0.0125,0.025){30}{\rule{0.3pt}{0.3pt}}
  \multiput(0.2,0)(0.0125,0.025){30}{\rule{0.3pt}{0.3pt}}
  \put(0,0.75){\line(0,-1){0.15}}
  \put(0.015,0.75){\line(0,-1){0.1}}
  \put(0.03,0.75){\line(0,-1){0.075}}
  \put(0.045,0.75){\line(0,-1){0.05}}
  \put(0.05,0.75){\line(0,-1){0.025}}
  \put(0.6,0){\line(0,1){0.15}}
  \put(0.585,0){\line(0,1){0.1}}
  \put(0.57,0){\line(0,1){0.075}}
  \put(0.555,0){\line(0,1){0.05}}
  \put(0.55,0){\line(0,1){0.025}}
  \end{picture}}
\newcommand{\Zbar}{\mathord{\!{\usebox{\zzzbar}}}}

\newcommand{\Z}{\Zbar}

\newcommand{\E}{\Ebar}
\begin{document}

\setlength{\textheight}{21cm}

\title{{\bf Entropy Production in Driven \\Spatially Extended Systems}}

\author{{\bf Christian
Maes}\thanks{Email: Christian.Maes@fys.kuleuven.ac.be }
\\ Instituut voor
Theoretische Fysica
\\ K.U.Leuven, B-3001 Leuven, Belgium.}

\maketitle

\begin{abstract} This is a short review of the statistical
mechanical definition of entropy production for systems composed of a
large number of interacting components.  Emphasis is on open systems
driven away from equilibrium where the entropy production can be
identified with a logarithmic ratio of microstate multiplicities of the
original macrostate with respect to the time-reversed state. A special
role is taken by Gibbs measures for the stationary spatio-temporal
distribution of trajectories. The mean entropy production is always
non-negative and it is zero only when the system is in equilibrium. The
fluctuations of the entropy production satisfy a symmetry first observed
 in \cite{ECM} and then derived in \cite{GC} for the phase space contraction
rate in a class of strongly chaotic dynamical systems.  Aspects of the
general framework are illustrated via a bulk driven diffusive lattice gas.
\end{abstract}
\vspace{3mm} \noindent

\section{Introduction}

The production of entropy in spatially extended systems was already
discussed, both at length and in depth, by the founding fathers of
thermodynamics and statistical mechanics. It is
instructive to divide the discussion in two (very unequal) parts.\\
First there is the situation of a perfectly closed (or truly isolated)
classical mechanical system undergoing a Hamiltonian dynamics and
evolving towards equilibrium from nonequilibrium initial conditions. That
is known as the approach or the convergence or the relaxation towards
equilibrium. The issue here is to understand the usual second law of
thermodynamics that associates an arrow of time to the macroscopic
behavior leading to equilibrium and to derive from the reversible
microscopic laws the irreversible kinetic and hydrodynamic equations
describing this evolution on the appropriate space-time scale.\\
Secondly, there are the many phenomena of dissipation in open driven
systems. Here one considers a stationary situation of a system or medium
in contact with several reservoirs that interact with the system. Think
of heat baths or particle reservoirs at different temperatures or
concentrations through which a nonequilibrium state is maintained in the
system or also of particle systems subject to an external driving field.
As a result, a stationary current is installed and entropy is produced.

The plan is that I start by recalling some of the main ingredients in the
qualitative understanding of macroscopic irreversibility for the first
scenario,
 return to equilibrium.  That has obtained most attention in the past and
 it is well understood.  On the other hand, for
 the nonequilibrium steady state scenario the literature is mostly
 limited to the close to equilibrium situation and
 Section \ref{sect:phen} must remind us of the
 phenomenology of entropy production.  Then, I get a chance to say something new when
  in Section \ref{sect:mul} I give
 a Boltzmann-like definition of entropy production in a simplified context.
 Formula (\ref{tata}) is most important. The general
framework is presented in Sections \ref{sect:gibbs} and \ref{sect:hyp}.
The application in the final Section \ref{sect:asep} is on the asymmetric
exclusion process.

The paper uses and studies aspects of thermodynamics, statistical
mechanics and probability theory with a variety of language and tools.
One application and source of inspiration is however not included, that
of the theory of smooth dynamical systems.  Yet, some of the work that I
will present here was started in \cite{M} (see also \cite{LS,K}) from
trying to understand the Gallavotti-Cohen results in \cite{GC} on the
fluctuations of the phase space contraction in dynamical systems with
very strong hyperbolicity assumptions
 and the `new theoretical ideas in nonequilibrium statistical
mechanics' \`a la Ruelle, see \cite{R}.  Some apology can be read in
Section \ref{sect:hyp}.\\
The present paper is, except for the Sections \ref{sect:mul} and
\ref{lftasep}, a short review and, by lack of space, the many details and
proofs are omitted.  More can be found in \cite{M}-\cite{MR} and I am
especially grateful to  Frank Redig for the joint work.

\section{Approach to equilibrium}

Here I only remind the reader about the problem of irreversibility in
closed Hamiltonian systems and about its resolution. There is nothing new
here (see e.g. \cite{G,L,J,B}) except perhaps for some measures of
emphasis.

What is the problem?  The microscopic dynamics is time-reversal invariant.
Why then do we always see individual macroscopic systems taking a
particular course in time, evolving towards equilibrium? What makes the
direction of time in
macroscopic behavior?\\
The resolution of this paradox has many sides and it can easily take you
away for quite some time.  Yet the beginning is easy and it says that
there is absolutely no problem, it is a matter of counting and you could
not expect otherwise: equilibrium is what you should expect. Here the
basic ingredient is that there is a huge difference in scales between the
microworld and the macroworld and that irreversibility belongs to the
macroworld.

\subsection{Boltzmann entropy}
Take a system of $N$ (large!) components $x(i), i=1,\ldots, N$, each
having $n$ possible values. A microstate $x= (x(i), i=1,\ldots, N)$ is
one of the $n^N$ possible configurations. The set of all these
configurations (or, when additional constraints are given, some subset of
this) is the microscopic phase space $\Omega$. From a microscopic
perspective each element of $\Omega$ is equivalent and we remain
indifferent in the estimates of plausibility: each $x$ has the same
probability for being realized.  From a macroscopic point of view, some
distinction can be made.  For this we make a choice of macroscopic
variables which are approximately additive (and, in presence of a
`natural' dynamics, are locally conserved). The macrovariable partitions
$\Omega$ in a number of subsets. Say for $\Omega=\{+1,-1\}^N$ and
macrovariable $X_N(x) \equiv \sum_i x(i)/N$ each of these subsets will
contain between 1 and, for $N$ very large, almost $2^N$ elements.  So we
all expect that randomly picking an element from $\Omega$ gives us a
configuration that `typically' has zero $X_N$ (with standard error
$1/\sqrt{N}$). In other words, while all microstates are equivalent, some
microstates are more typical
from a macroscopic point of view.\\
 To connect the microscopic world with
macroscopic behavior Boltzmann introduced the entropy
\[
S(X)\equiv \log W(X)
\]
where $W(X)$ is the number of microstates $x$ for which $X_N(x)=X$.  The
logarithm is very convenient and makes $S(X)$ extensive.  In the same way
(and by some abuse of notation), we speak about the entropy $S$ of a
microstate $x$:
\[
S(x) \equiv S(X_N(x)).
\]
Of course, the phase space for realistic systems is a bit more
complicated.  We should be speaking about a (classical) Hamiltonian
dynamics for $N$ point particles enclosed in a finite box with $x$
corresponding to the positions and momenta of all the particles.  The
phase space $\Omega$ is now the collection of all these microstates
constrained to given values of certain macrovariables $Y_N$. As we have a
closed system with energy conservation, we can keep in mind that the
total energy $Y_N(x)=E$ is thus fixed. We are interested in the evolution
of a second class of macrovariables $X_N$. Let us think here about the
macroscopic variable that corresponds to the spatial density profile, say
the number of particles to the left of our box. Yet, for this more
complicated phase space, more or less the same counting procedure can be
followed. In this same language, microstates have a larger entropy when
they correspond to a macroscopic value that can be microscopically
realized in more ways. Now it is important to get a feeling of the huge
differences that can arise when the system is large. For air, under
normal circumstances, there are about $10^{25}$ molecules in a box of 1
$m^3$.  If we divide this box in pieces of 1 $cm^3$ we get $10^{60 000
000 000 000 000 000 000 000}$ possible arrangements that would correspond
to a homogeneous density profile (about the same number of particles in
each piece) while only $10^6$ possibilities of having all of them in just
one such piece of the box. When considering these huge numbers and
reasoning about plausibilities just as we do in everyday life, it appears
that only a great conspiracy of initial conditions or dynamics can lead
to an effective decrease in entropy: that is, we expect that $S(x_t)$
grows as $x_t$ follows the microscopic trajectory and equilibrium
corresponds to maximal entropy.
 The law of increase of entropy is a statistical law,
but one with a `moral certainty.'\\
  It must be added that larger entropy does
not necessarily mean less ordered or more spatially homogeneous.  It
depends on the type of interaction and/or the value of the energy that is
fixed.  It suffices to think of gravity (where clustering of mass is
typical) or of molecular interactions at low energy leading to ordered
structures.

\subsection{Initial conditions}
Note that the counting argument above is symmetric in time, valid as such
for prediction as well as for for retrodiction.  So the real surprise is
to see nonequilibrium initial conditions.  When trying to see the origin
of these special low entropy initial conditions we soon get in a chain of
arguments leading to cosmological questions.  Here input from the
standard model of physical cosmology is necessary.  The upshot is
summarized by Richard Feynman:
``it is necessary to add to the physical laws the hypothesis that in the past the universe was
more ordered, in the technical sense, than it is today...to make an
understanding of the irreversibility'' or, much earlier, by Ludwig
Boltzmann:
``That in nature the transition
from a probable to an improbable state does not take place as often as the
converse, can be explained by assuming a very improbable initial state of
the entire universe surrounding us.  This is a reasonable assumption to
make, since it enables us to explain the facts of experience, and one
should not expect to be able to deduce it from anything more
fundamental.'' I refer to the pleasant Chapter 7 in \cite{Penrose} for
some specific analysis.

\section{Phenomenology of steady state entropy
production}\label{sect:phen} The definition of entropy production for
nonequilibrium steady states cannot be completely arbitrary.  Standard
treatments are however largely restricted to close to equilibrium
phenomenology.

Entropy production appears in the close to equilibrium thermodynamics of
steady state irreversible processes as the product of thermodynamic
fluxes and thermodynamic forces. The forces are produced by gradients in
intensive variables, the so called affinities that are maintained
throughout.  One thinks of gradients in chemical or electrostatic
potential or in temperature. The fluxes are currents in the conjugate
extensive variables. In linear response, they are linear combinations of
the affinities.  In this way, linear transport coefficients are defined
that are functions of the intensive variables that locally
characterize the state of the system.\\
As an example take a cylindric solid material through which a stationary
heat current $J_Q$ is maintained by coupling the material at its right
and left ends to reservoirs at temperature $T_r$ and $T_{\ell}$
respectively.  Assuming a linear relation between the affinity $\nabla
1/T$ and the current $J_Q$ with a linear response coefficient $L_Q$, one
arrives at Fourier's law
\[
J_Q = L_Q \nabla (\frac 1{T}) = - h \nabla T
\]
where $h \equiv L_Q/T^2$ is the heat conductivity. Here is expressed the
empirical finding (close to equilibrium) of the proportionality of the
heat current with the temperature gradient.  The entropy production in
the material is then
\[
\dot{S}= J_Q \nabla \frac 1{T} = h T^2 [\nabla(\frac 1{T})]^2
\]
Of course, the (close to equilibrium) entropy of the material remains
constant (its macrostate is unchanged) and all produced entropy is
carried away by the entropy current to be poured into the reservoirs. (The
relation between entropy current and entropy production is established in
the so called entropy balance equation.)  If we integrate the entropy
production over the material, we see that the entropy of the reservoirs is
changed by
\[
\frac{dS}{dt} = a J_E (\frac 1{T_r} - \frac 1{T_\ell})
\]
where $a$ is the cross-section area of the cylinder.  To maintain the
steady state we will have to carry away this extra entropy outside the
coupled system.  I refer to \cite{Bal} for the standard treatment and to
\cite{BL,Eck} for more recent studies.

\section{Multiplicity under constraints}\label{sect:mul}
An elementary mathematical clarification of the connection between
Boltzmann entropy and thermodynamic equilibrium entropy goes by
considering the multiplicity of microstates for a particular
macroscopic observation, (see also \cite{Geo}).\\
Take $N$ particles each of which can be in a certain phase space cell $i$
(for example having energy $E_i), i=1,\ldots,n$. In total, there are
$n^N$ microstates.  The number of such microstates with $m_1$ particles
in cell $i=1$, ... and $m_n$ particles in cell $i=n$ ($m_1 + \ldots +
m_n= N$) is given by the multinomial coefficient
\[
W_N(m_1,m_2,\ldots,m_n) = \frac{N!}{m_1!\ldots m_n!}
\]
With proportions $p_i \equiv m_i/N$ and via Stirling's formula,
\begin{equation}\label{stir}
\lim_N \frac 1{N} \ln W_N(p_1N,\ldots,p_nN)=-\sum_{i=1}^n p_i \ln p_i
\end{equation}
where the right hand side is the Shannon entropy of the probability
measure $(p_i)$. Imagine now as further constraint that
\[
\sum_{i=1}^n p_i E_i = E
\]
Then, the multiplicities or the Shannon entropy of (\ref{stir}) is
maximal when
\[
p_i = \frac{e^{-\beta E_i}}{Z_\beta}
\]
with $\beta=\beta(E)$ found from the constraint. This maximal entropy is
the equilibrium or Gibbs' entropy
\[
S(E) = \ln Z_\beta + \beta E
\]
We can now move $E$ to a new value $E+dE$ while supposing that the energy
levels $E_i$ do not change. Obviously, the maximal entropy does change
and it is doing so according to
\[
dS = \beta dE
\]
which is Clausius' formula $dS= \delta Q/T$ for the change in equilibrium
entropy from a heat transfer $\delta Q$ at reservoir temperature
$T=1/\beta$ for fixed volume.

I suggest a very similar scheme for entropy production.  I sketch it here
without explicit reference to dynamics just to emphasize, in this highly
simplified setting, the structure of the definition that later, in fuller
glory, will follow.  (In brackets I will announce the analogue.)\\
Consider again $N$ variables $\sigma(i), i=1\ldots,N$ with, for
simplicity, $\sigma(i) =0,1$, i.e., $n=2$ in the above (this should be
thought of as the pathspace, the space of all microscopic space-time
trajectories.  Thus $N$
 is the number of particles times the number of times
they are observed;  $\sigma(i)=1$
 indicates a right-moving particle at that time while
$\sigma(i)=0$ indicates left-moving). Let us select the first $M\leq N$
of these variables and let $(\eta(i), i=1,\ldots,M)$ be a fixed
configuration on them. (That is, we select a particular history for a
given space-time volume.) That of course constrains the number $m_1$ of
all variables that are in the first microstate ($1$):
\[
N - M +  \sum_{i=1}^M \eta(i)  \geq m_1 \geq \sum_{i=1}^M \eta(i)
\]
Let $W_M^{(N)}(\eta,m)$ be the number of microstates $\sigma\in
\{0,1\}^N$ so that $\sigma(i) = \eta(i)$ for  $i=1,\ldots,M$ and so that
\[
\sum_{i=1}^N \sigma(i) = mN
\]
(That will correspond to a given macroscopic value for some current.)
Consider the involution $\eta(i) \rightarrow \bar{\eta}(i) \equiv
1-\eta(i), i=1,\ldots,M$ (time-reversal will take that role later). I
suggest now to inspect the logarithmic ratio
\begin{equation}\label{tata}
\dot{S}_M^{(N)}(\eta,m) \equiv \ln
\frac{W_M^{(N)}(\eta,m)}{W_M^{(N)}(\bar{\eta},m)}
\end{equation}
(That will be the entropy production.) Evaluating this for large $N$,
while fixing all the rest, gives
\begin{equation}\label{je}
\dot{S}_M(\eta,m) \equiv \lim_N \dot{S}_M^{(N)}(\eta,m) = \sum_{i=1}^M
[\eta(i) - \bar{\eta}(i)] \ln \frac{m}{1-m}
\end{equation}
(Observe that for $\lambda \equiv \ln m/(1-m)$ indeed $\exp \lambda/(1 +
\exp \lambda) = m$ so that (\ref{je}) is the product of the field
$\lambda \neq 0$ for $m\neq 1/2 (=$ the equilibrium value) and the
variable current or flux in the space-time volume $M$.) If we now also
let $M$ grow very large, we will have that
\[
\sum_{i=1}^M \eta(i) \approx mM, \sum_{i=1}^M [\eta(i) -
\bar{\eta}(i)]\approx (2m-1)M
\]
and hence, $\dot{S}_M(\eta,m)/M$ converges to the  relative entropy
\[
\dot{S}(m) = (2m-1)\ln\frac{m}{1-m} = m \ln \frac{m}{1-m} + (1-m) \ln
\frac{1-m}{m}
\]
(That is, the mean entropy production is positive and it equals the
relative entropy between the original distribution and its time-reversal.)

\section{Gibbs measures with an involution}\label{sect:gibbs}

The following must be considered as a program, the ideal context that we
should keep in mind for more concrete realizations.

I write $K$ for a finite set and take $\Omega \equiv K^{\Z^{d+1}}$. This
$(d+1)-$dimensional configuration space announces that $\Omega$ plays the
role of pathspace; its elements $\sigma \equiv (\sigma_t(i), (t,i)\in \Z
\times \Z^d)$ are space-time trajectories with values $\sigma_t(i) \in K$
at the space-time point $(t,i)$.  Consider regular space-time cubes
$V_{T,L} \equiv \{(t,i)\in \Z^{d+1}: |t| \leq T, |i|\leq L\}$ centered
around the origin, where $T$ stipulates the temporal and $L$ the spatial
extension. Let $\Theta_{T,L}$ be the involution on $\Omega$ that reverses
the time:
\begin{eqnarray}
(\Theta_{T,L}\sigma)_t(i) &\equiv& \sigma_{-t}(i) \mbox{ if } (t,i)\in
V_{T,L},\\ \nonumber & \equiv & \sigma_t(i) \mbox{ otherwise}
\end{eqnarray}
(I could also have included a kinematical time-reversal $\pi$ as
involution on $K$ and then write $(\Theta_{T,L}^\pi \sigma)_t(i) \equiv
\pi(\sigma_{-t}(i))$ for $(t,i)\in V_{T,L}$ but, for simplicity, I will
stick here to the choice $\pi=$identity.) Given a local function $f$ on
$\Omega$ it is possible to find large enough $T_o, L_o$ so that for all
$T\geq T_o, L\geq L_o, f(\sigma) = f(\sigma_t(i), (t,i) \in V_{T,L})$.
There is therefore no ambiguity in writing $f\Theta$ for the new
(time-reversed) function and similarly, for a probability measure $\mu$
on $\Omega$, to write $\mu\Theta$ for the new probability measure with
expectations $\mu\Theta(f) \equiv \mu(f\Theta)$.\\
Next, consider a translation invariant space-time interaction potential
$U= (U_A)$ parametrized by the finite subsets $A$ of $\Z^{d+1}$.  Each
$U_A$ is a function of the variables $\sigma_t(i), (t,i)\in A$, and let
 me assume, just for convenience, that $U_A\equiv 0$ whenever the diameter of the set $A$
 is larger than a finite radius. I define
the associated entropy production in a finite set $\Lambda$ as
\begin{equation}\label{en}
\dot{S}_\Lambda \equiv \dot{S}_\Lambda^{(U)} \equiv \sum_{A \subset
\Lambda} [U_A\Theta - U_A]
\end{equation}
Clearly, $\dot{S}_\Lambda$ is asymmetric under time-reversal $\Theta$ and
it is measurable from the values of the variables in the set $\Lambda$.
Definition (\ref{en}) is the analogue of (\ref{tata}). In all realistic
realizations of this definition, $\dot{S}_\Lambda$ can be interpreted as
a sum over products of fields and currents. The reader is probably
waiting to see some dynamics and it may seem strange to speak about such
an entropy production but I only deal with the general framework here and
one illustration is contained
in Section \ref{sect:asep}. \\
Let $\mu$ be a translation-invariant probability measure on $\Omega$.  I
define the mean entropy production (MEP) in $\mu$ as the expectation
\begin{equation}\label{mep}
\mbox{MEP}(U,\mu)\equiv \lim_{\Lambda} \frac 1{|\Lambda|}
\mu(\dot{S}_\Lambda)
\end{equation}
Limits are understood in the sense of increasing cubes $\Lambda=V_{T,L}$.
It is important to realize that you do not in fact need to take the
$\mu-$average in the above but for a fixed (large enough) $T,
\dot{S}/|\Lambda|$ will become $\mu-$almost surely equal to the MEP if
$\mu$ is ergodic with
respect to spatial translations.\\
Let $\mu$ be a translation invariant Gibbs measure for the potential
$(U_A)$.  It has an entropy density $s(\mu)$ that equals the entropy
density $s(\mu\Theta)$ of the time-reversed Gibbs distribution. Yet,
$\mu$ and $\mu\Theta$ need not be equal; they can be discriminated via
their relative entropy density $s(\mu|\mu\Theta) = s(\mu\Theta|\mu)$ (see
\cite{Geo}).

\noindent {\bf Theorem 1 (MEP)} Let $\mu$ be a translation invariant
Gibbs measure for the potential $(U_A)$. Then
\begin{equation}\label{meprel}
\mbox{MEP}(U,\mu) = s(\mu|\mu\Theta) \geq 0
\end{equation}
with equality if and only if the potentials $U$ and $U\Theta$ are
physically equivalent.

\noindent {\bf Remarks}:
\begin{itemize}
\item
The identification of the MEP with the relative entropy density was
announced at the very end of Section \ref{sect:mul}.
\item
There can be no spontaneous breaking of time-reversal symmetry: If the
MEP is zero, then $\mu$ and $\mu\Theta$ must be Gibbs measures for the
same potential, which means that $(U_A)$ and $(U_A\Theta)$ must be
physically equivalent. This property of `no current without heat'  is
established for infinite interacting particle systems in
\cite{MR,MRV1,MRV2}.
\end{itemize}

I now present the local (in space) fluctuation theorem (LFT) for Gibbs
measures with an involution.  Take $T$ and $L$ large and fix the volume
$V \equiv V_{T,L}$ and a subset $\Lambda \equiv V_{T,L'}$ with $L' <<
L$.  For a function $f$ of the variables in $V$, define the expectation
\[
\E_V(f) \equiv \frac 1{Z_V} \sum_{\sigma \in K^V} f(\sigma)
e^{-\sum_{A\subset V} U_A(\sigma)}
\]
with $Z_V$ the normalizing partition function. There is no need for
translation invariance here and the potential $(U_A)$ is allowed to
contain time-reversal invariant hard core interactions. I write
\[
Z_{V\setminus\Lambda}(\sigma_\Lambda) \equiv \sum_{\sigma_{ \Lambda^c}
\in K^{V\setminus\Lambda}} e^{-\sum_{A\subset V, A\cap \Lambda^c\neq
\emptyset} U_A(\sigma_{\Lambda^c}\sigma_\Lambda)}
\]
for the partition function in $\Lambda^c \equiv V\setminus\Lambda$ with
boundary condition $\sigma_\Lambda$ in $\Lambda$.  Put
\[
F_\Lambda^V(\sigma_\Lambda) \equiv \ln
\frac{Z_{V\setminus\Lambda}(\Theta\sigma_\Lambda)}
{Z_{V\setminus\Lambda}(\sigma_\Lambda)}
\]
which, for local interactions, depends on the variables in (the interior
boundary $\partial \Lambda$
of) $\Lambda$.\\
Put $R_\Lambda^V \equiv \dot{S}_\Lambda - F_\Lambda^V$.\\
\noindent {\bf Theorem 2 (LFT)}
\begin{itemize}
\item
For every function $G$
\begin{equation}\label{gc1}
\E_V[G(-R_\Lambda^V)] = \E_V[G(R_\Lambda^V)e^{-R_\Lambda^V}]
\end{equation}
\begin{equation}\label{gc}
\E_V[G(-\dot{S}_\Lambda)] = \E_V[G(\dot{S}_\Lambda)e^{-\dot{S}_\Lambda +
F_\Lambda^V}]
\end{equation}
\item
For every family  $(\lambda_A)$ of complex numbers
\begin{equation}\label{genf}
\E_V [e^{-\sum_{A\subset\Lambda} \lambda_A [U_A\Theta-U_A]}] = \E_V
[e^{-\sum_{A\subset\Lambda} (1-\lambda_A) [U_A\Theta-U_A]}
e^{F_\Lambda^V}]
\end{equation}
\end{itemize}

\noindent{\bf Remarks}:
\begin{itemize}
\item
I take it understood that, for sufficiently local interactions, the
potential $F_\Lambda^V$ is of the order of the boundary of $\Lambda$.
Therefore, asymptotically in logarithmic sense, (\ref{gc}) leads directly
to the symmetry (\ref{gc1}) for the distribution of the entropy
production $\dot{S}_\Lambda$ itself, a symmetry first uncovered in
\cite{ECM,GC} for the large deviation rate function of the phase space
contraction in the SRB state of reversible dissipative mixing Anosov
diffeomorphisms.  One should however not take for granted that
$F_\Lambda^V$ is uniformly bounded by $|\partial \Lambda|$.  In many of
the models to which one wishes to apply the above scheme, one really
needs to prove
 that it concerns here just a `boundary term.'  One origin of problems can be that time is
 taken
continuous and an unbounded number of changes can happen in any finite
interval.  This is less of a problem for jump processes where  one uses
that the Poisson process has all exponential moments but when working on
non-compact phase spaces, the `boundary term' $F_\Lambda^V$ can possibly
have non-existing exponential moments and thus really can change the
large deviation rate function.
\item
The identity (\ref{genf}) generates, by suitable differentiation,
equalities between correlation functions.  A discussion with an
application to the Onsager reciprocity relations is contained in
\cite{M}.  I like to compare (\ref{genf}) with the Ward identities as we
know them from quantum field theory; the mathematical origin is very
similar and non-perturbative, liberating us from close to equilibrium
assumptions.
\end{itemize}

\section{Gibbsian hypothesis}\label{sect:hyp}

The ambition in the above framework for the study of nonequilibrium
driven systems is to use the standard Gibbs formalism and in particular
its fluctuation theory in the `current'-ensemble. The mathematics will
therefore not deviate significantly from
what is e.g. found in the contributions \cite{denH,Var,Geo}.\\
  A frequently asked question  is
where these Gibbs measures come from or how they should be connected with
a dynamics.  Here two short answers.

\subsection{Pathspace measure construction}
The first answer is illustrated below in Section \ref{sect:asep} and many
more examples can be found in \cite{M}-\cite{MR} and \cite{LS,K, Eck}.
One constructs the Gibbs measure explicitly as the pathspace measure of a
stationary process.  Most easy is the case of stochastic dynamics.  It is
not really necessary to realize the pathspace measure as a {\it bona
fide} Gibbs measure as we know it say for lattice spin systems; what is
needed is to understand how the pathspace measure is governed by a
space-time action which is approximately local and additive in
space-time.  Technically, this is a matter of setting up the appropriate
Girsanov formula and I refer to the standard treatments in \cite{LS1,Bre}.
  Interestingly enough, this method of constructing the process via
a Gibbsian space-time distribution has also lead to new existence and
uniqueness results for
classes of diffusion processes, see e.g. \cite{sylvie}.\\
Deterministic dynamics are here more of a problem but, for example, we
have learnt from \cite{G1}-\cite{GC} how the Gibbsian structure can be
exploited on the level of the symbolic dynamics for sufficiently strongly
chaotic dynamics.

\subsection{Maximum entropy principle}
The first answer has the advantage of being explicit and directly
connected to model dynamics that one may have in mind. I believe however
that the second answer is more to the point. The reason to use Gibbs
measures here is exactly the same as for using Gibbs measures in usual
classical or quantum equilibrium statistical mechanics.  All that changes
is the type of ensemble because we must work with the currents as
macro-observables.  To do this on space-time is the only price that must
be paid but it is a necessary one. Gibbs measures appear then as solutions
of the maximum entropy principle but that need not be restricted to
equilibrium conditions.

\section{Asymmetric exclusion process}\label{sect:asep}
I illustrate the previous generalities using one concrete model, that of a
 bulk driven diffusive lattice gas.
 First about the mean entropy production, to check that it coincides with
 what one should expect.  Secondly, about the local entropy production fluctuations.

\subsection{MEP for ASEP}\label{mepasep}  I start with the asymmetric exclusion
process (ASEP) on a one-dimensional ring $\{1,2,\ldots,\ell\}$ (periodic
boundary conditions). Each site $i$ of the ring is either occupied by a
particle (denoted by $\eta(i)=1$) or is empty ($\eta(i)=0$).  The
particle-configuration $\eta$ is subject to a Markovian particle
conserving asymmetric hopping dynamics with rates
\begin{equation}
c(i,i+1,\eta) = \eta(i)(1-\eta(i+1)) \frac{e^{E/2}}{2}
 + \eta(i+1)(1-\eta(i)) \frac{e^{-E/2}}{2}
\end{equation}
for changing the configuration from $\eta$ to $\eta^{i,i+1}$ obtained by
exchanging the occupations at sites $i$ and $i+1$.  In words, particles
hop to nearest neighbor sites when there is a vacancy at a rate that
depends on the direction. $E$ is now an external driving field. We can
easily obtain the space-time interaction from the standard Girsanov
formula for Markov chains.  The entropy production is then the relative
action under time reversal and its expectation in a steady state is the
MEP.\\
The product measure $\rho_u$ with uniform density $u\in [0,1]$ is a
stationary (non-reversible) measure for this dynamics. If we now consider
a trajectory $(\eta_t, t\in [-T,T])$ of the stationary process in which
at a certain time, when the configuration is $\eta$, a particle hops from
site $i$ to $i+1$, then the time-reversed trajectory shows a particle
jumping from $i+1$ to $i$.  The contribution of this event to the entropy
production is therefore
\begin{equation}\label{asym}
\ln c(i,i+1,\eta) -\ln c(i,i+1,\eta^{i,i+1}) = E[\eta(i)(1-\eta(i+1))-
\eta(i+1)(1-\eta(i))]
\end{equation}
This jump in the trajectory itself happens with a rate $c(i,i+1,\eta)$
and therefore the mean entropy production equals
\begin{eqnarray}
\label{hydro} \mbox{MEP}(u,E) &=& \int c(0,1,\eta) \ln
\frac{c(0,1,\eta)}{c(0,1,\eta^{01})}
 \rho_u (d\eta) \nonumber \\
 &=& E \ u \  (1-u) \ \sinh(\frac{E}{2})
\end{eqnarray}
That is correct: the MEP is the product of the field $E$ with the current
$j(u,E)$ where the current $j(u,E)$ is the expected net number of
particles passing through a given bond:
\begin{eqnarray}\nonumber
j(u,E) & \equiv &\int \rho_u(d\eta) c(i,i+1,\eta) [\eta(i)(1-\eta(i+1))-
\eta(i+1)(1-\eta(i))] \\\nonumber &=&  u(1-u) \sinh(\frac{E}{2})
\end{eqnarray}
In quadratic approximation (that is close to equilibrium) $j(u,E) \approx
u(1-u) E/2$ and,
\begin{equation}
\mbox{MEP}(u,E) \approx \frac{j(u,E)^2}{h_c}
\end{equation}
which is the dissipated heat through a conductor in an electric field $E$
with Ohmic conductivity $h_c \equiv u(1-u)/2 = \rho_u([\xi(0)(1-\xi(1)) -
\xi(1)(1-\xi(0))]^2)/4$ given in terms of the variance of the microscopic
current (at $E=0$).

\subsection{LFT for ASEP}\label{lftasep}

I take a $(2+1)-$dimensional set-up.  There are (spatial) squares $V_0
\equiv [-L,L]^2 \cap \Z^2$ and $\Lambda_0 \equiv [-L',L']^2 \cap \Z^2$
with $L' < L$ large, and a continuous time interval $[-T,T]$ in which we
observe the ASEP with the external field $E$ in the horizontal direction.
That is the 2-dimensional analogue of \ref{mepasep} but now on $V_0$ with
periodic boundary conditions and hopping rates
\[
c(i,j,\eta)\equiv \frac{e^{E/2}}{2}\eta(i)(1-\eta(j)) +
\frac{e^{-E/2}}{2}\eta(j)(1-\eta(i))
\]
for a horizontal bond $\langle i j=i+e_1\rangle$ with $e_1$ the unit
vector in the positive horizontal direction, and
\[
c(i,j,\eta)\equiv \frac 1{2}[\eta(i)(1-\eta(j)) + \eta(j)(1-\eta(i))]
\]
for a vertical bond $\langle i j=i\pm e_2\rangle$.  As stationary measure
I take again $\rho_u$ (Bernoulli with density $u$) and $(\eta_s(i), s\in
[-T,T],
i\in V_0)$ denotes the stationary process.\\
I show what becomes of relations (\ref{gc1}) and (\ref{gc}).  Let
$\E_V^E[\cdot]$ denote the expectation with respect to the process in $V
\equiv [-T,T] \times V_0$ with (pathspace) law $P_V^E$.
  For a function $f$ measurable from $\Lambda \equiv [-T,T] \times
\Lambda_0$,
\begin{eqnarray}\label{defrn}
\E_V^E[f \Theta_{T,L'}] &=& \E_V^E[f \Theta_{T,L}]= \E_V^E[f \frac{d(P_V^E
\Theta_{T,L})}{dP_V^E}]\\ \nonumber &=& \E_V^E[f
\frac{dP_V^{-E}|_\Lambda}{dP_V^E|_\Lambda}]
\end{eqnarray}
so that, with
\[
R_\Lambda^V = \ln \frac{dP_V^{E}|_\Lambda}{dP_V^{-E}|_\Lambda}
\]
and $f= G(R_\Lambda^V)$, the identity (\ref{defrn}) is just (\ref{gc1}).
$R_\Lambda^V$ is computed from a Girsanov formula for the non-Markovian
point proces $dP_V^{E}|_\Lambda$ and for this I first need to identify
the intensities (see \cite{LS1,Bre}).  In order to have a Gibbsian
structure (allowing me to pass to (\ref{gc})) these intensities must be
the same in the bulk of $\Lambda$ as they were in the bulk of $V$.  That
is easy to verify from considering the conditional expectation of a
function $g$ in $\Lambda_0$ at time $t+\delta$ given the past history in
$\Lambda_0$:
\begin{eqnarray}
&&\E_V^E[g(\eta_{t+\delta}(i),i\in \Lambda_0)|\eta_s(i), s\in [-T,t],i\in
\Lambda_0] =\\ \nonumber &&\E_V^E[\E_V^E[g(\eta_{t+\delta}(i),i\in
\Lambda_0)|\eta_s(i), s\in [-T,t],i\in V_0]|\eta_s(i), s\in [-T,t],i\in
\Lambda_0]
\end{eqnarray}
The conditional expectation inside is explicit from the Markov process in
$V_0$:
\begin{eqnarray}
&&\E_V^E[g(\eta_{t+\delta}(i),i\in \Lambda_0)|\eta_s(i), s\in [-T,t],i\in
V_0] \\ \nonumber && = g(\eta_{t}(i),i\in \Lambda_0) + \delta
L_V^Eg(\eta_t(i), i\in V_0) + O(\delta^2)
\end{eqnarray}
with $L_V^E$ the Markov generator of the process.  Since $L_V^E$ is a sum
over all bonds with local rates, we see that the process restricted to
$\Lambda_0$ has the same rates except for the boundary of $\Lambda_0$
where a birth and death process is added.  As a result the Girsanov
formula for $R_\Lambda^V$ is indeed
\[
R_\Lambda^V = \dot{S}_\Lambda - F_\Lambda^V
\]
measurable in $\Lambda$ with
\begin{equation}\label{enasep}
\dot{S}_\Lambda(\eta_\Lambda) = E \int_{-T}^T \sum_{i,i+e_1\in \Lambda_0}
[\eta_t(i)(1 -\eta_t(i+e_1)) - \eta_t(i+e_1)(1-\eta_t(i))] dN_i^1(t)
\end{equation}
where $N_i^1(t)$ is the number of jumps between $i$ and $i+e_1$ up to time
$t$, and
\begin{eqnarray}\label{bdasep}
F_\Lambda^V(\eta_\Lambda) &=& \sum_{i\in \partial \Lambda}\sum_{b_i}
\int_{-T}^T
[\ln\frac{\kappa^E_{b_i}(\eta,t)}{\kappa^{-E}_{b_i}(\eta,t)}\\ \nonumber
&&+ \eta_t(i) \ln\frac{\lambda^E_{b_i}(\eta,t)\kappa^{-E}_{b_i}(\eta,t)}
{\lambda^{-E}_{b_i}(\eta,t)\kappa^{E}_{b_i}(\eta,t)}] dN_{b_i}(t)
\end{eqnarray}
where the second sum is over all bonds $b_i$ starting at site $i\in
\Lambda_0$ with the other end $j\in \Lambda_0^c$ and, with $b_i=\langle
ij\rangle$,
\[
\lambda^E_{b_i}(\eta,t) \equiv e^{E_j/2}\E_V^E[1-\eta_t(j)|\eta_s(k),
k\in \Lambda_0, s\in [-T,t]]
\]
and
\[
\kappa^E_{b_i}(\eta,t) \equiv e^{E_j/2}\E_V^E[\eta_t(j)|\eta_s(k), k\in
\Lambda_0, s\in [-T,t]]
\]
for $E_j=\pm E$ if $j=i\pm e_1$ and $E_j=0$ if $j=i\pm e_2$. The
expression (\ref{enasep}) is the entropy production, that is field times
current in $\Lambda$, and (\ref{bdasep}) is the boundary term.  That
establishes (\ref{gc}).

\end{document}